\documentclass[%
superscriptaddress,
 amsmath,amssymb,
 aps,
 twocolumn 
 ]{revtex4-2}

\usepackage{graphicx}% Include figure files
\usepackage{dcolumn}% Align table columns on decimal point
\usepackage{bm}% bold math
\usepackage{xcolor}
\usepackage[caption=false]{subfig}
\usepackage{xfrac}
\usepackage{blindtext}
\usepackage{times}
\usepackage{comment}
\usepackage{hyperref}
\usepackage{etoolbox}

\definecolor{darkblue}{HTML}{004D6B}
\definecolor{darkred}{HTML}{8c1515}
\hypersetup{
	colorlinks=true,
	urlcolor=darkblue,
	citecolor=darkred,
	linkcolor=darkblue,
	breaklinks
}

\begin{document}

\title{Photonic Qubit Gates via 1D Scattering from an Array of Two-Level Emitters}

\author{Evangelos Varvelis}
\email[Corresponding author: ]{evangelos.varvelis@uni-ulm.de}
\affiliation{Institute for complex quantum systems, University of Ulm, 89069 Ulm, Germany}

\author{Joachim Ankerhold}
\affiliation{Institute for complex quantum systems, University of Ulm, 89069 Ulm, Germany}

\date{\today}

\begin{abstract}
Photonic quantum computing offers a promising platform for quantum information processing, benefiting from the long coherence times of photons and their ease of manipulation. This paper presents a scheme for implementing a deterministic phase gate for dual-rail number encoded photonic qubits, leveraging a standard  1D waveguide coupled to an array of two-level emitters (TLE). Using a transfer matrix approach, we develop a protocol for deterministic phase gate operation, demonstrating its robustness against non-waveguide mode coupling and disorder. Finally, we relax the idealized assumption of monochromatic light, considering finite-bandwidth pulses. Despite these realistic considerations, our results indicate high fidelity for the proposed phase gate protocol. Finally we will discuss two qubit operations.
\end{abstract}

\maketitle

\section{Introduction}

Quantum computation using photons has gained significant momentum in recent years, driven by various technological advancements \cite{PhotonicComputingReview1,PhotonicComputingReview2}. A key feature of this approach is the long coherence time of photons, with state-of-the-art superconducting cavities enabling single-photon qubits to maintain coherence for up to 34 milliseconds \cite{PhotonCoherenceTime1,PhotonCoherenceTime2}. This stability is crucial for reliable quantum information processing. Moreover, photons are relatively easy to manipulate, facilitating the straightforward implementation of quantum gates and operations at the single-qubit level. Furthermore, photonic qubits represent a primary platform for quantum cryptography, allowing for secure communication methods such as quantum key distribution (QKD) \cite{QuantumCrypto1,QuantumCrypto2,QuantumCrypto3} at the speed of light, and they have demonstrated quantum supremacy in state of the art devices \cite{ChineseBosonSapmling}.

In order to operate a quantum system as a quantum computer, a universal gate set comprising both one- and two-qubit gates is essential. While the development of two-qubit gates using photons continues to face numerous challenges, significant advancements have been made in this area through the use of nonlinear elements \cite{PhotonicCPhaseGate} and linear elements combined with projective measurements \cite{KLMProtocol}. However, these approaches often introduce additional technical overhead, rely on probabilistic gates, or present other difficulties in experimental controllability. In this paper, we will concentrate on single-qubit operations, which are relevant in their own right \cite{SinglePhotonApplications}, with an emphasis on a single-qubit phase gate which is relevant for quantum cryptography \cite{QuantumCryptoApp}. Additionally, we also discuss the implications of our employed formalism for two-photon operations.

The main inspiration for our work has been an experimental setup \cite{StuttgartDevice,StuttgartDevice2} consisting of a silicon slot waveguide filled with dense thermal rubidium vapors, which enhances light-induced interactions between atoms through the Purcell effect. Atoms are excited using a resonant pump laser, while their interactions are probed with a second laser. This system allows for dynamic control of the nonlinearity at the few-photon level by tuning the interaction strength via the intensity of the driving field, thereby enabling the manipulation of photonic states. Furthermore, its ability to operate at room temperature presents a significant advantage over many quantum computing platforms that require cryogenic conditions, making it more suitable for integration into existing technologies. Additionally, the ability to enhance or suppress photon emission into selected modes opens new avenues for exploring topological \cite{StuttgartTopological} and chiral \cite{StuttgartChiral} quantum systems, and it can also serve as a memory using the method described in \cite{StuttgartMemory}.

In this work, we propose a scheme for realizing a passive, deterministic photonic phase gate using a one-dimensional array of two-level emitters (TLEs). While related architectures have been explored in previous studies \cite{Sorensen2022,Tudela2025}, many rely on chiral light-matter interactions—an approach that is difficult to implement in experimental platforms based on atomic vapors. Here, we show that high-fidelity phase gates can be achieved without relying on chiral transport, and that our protocol remains robust in the presence of photon loss and realistic forms of disorder. Motivated by the experimental setup described in \cite{StuttgartDevice}, we adopt a theoretical framework based on the rotating wave approximation of the multimode Dicke model, which enables generalization to other physical platforms. We begin by introducing this model and outlining our hybrid approach that combines the input-output formalism with the transfer matrix method. We then demonstrate how a periodic array of TLEs can implement a phase gate for dual-rail, number-encoded photonic qubits. We assess the resilience of the protocol to decoherence from non-waveguide losses, and subsequently extend the analysis to disordered arrays, relaxing the assumption of perfect periodicity. Our results indicate that the gate performance remains robust even in this more realistic regime. Furthermore, we analyze the scattering of finite-bandwidth photon pulses and show that for bandwidths narrow compared to the emitter linewidth, our protocol remains valid without modification. We conclude by discussing the implications of our formalism for the two-photon scattering regime, regarding its potential for generating photon-photon interactions via a single TLE.

We begin by introducing the model employed in this work, namely the rotating wave approximation of the multimode Dicke model \cite{DickeModel,DickeMultimodeModel}. The Hamiltonian of this model consists of three terms 
\begin{equation}\label{FullHamiltonian}
    H = H_0 + H_1 + H_\text{int}
\end{equation}
the first term describes the energy of freely propagating photons in the waveguide.
\begin{equation}
    H_0 =  \hbar\int d\omega\ \omega\left(\ell_\omega^{\dagger} \ell_\omega + r_\omega^{\dagger} r_\omega\right),\label{WaveguideHamiltonian}
\end{equation}
where the photonic modes have been separated in two species: left-propagating modes created by the operator $\ell_\omega^{\dagger}$, and right-propagating modes created by the operator $r_\omega^{\dagger}$. Unless otherwise stated, integrals appearing throughout this paper without explicit limits are taken over the entire real axis. The reason we have extended the frequency integration limits to the negative axis is because of the assumption of a narrow-band photon source \cite{NegativeFrequencies}. The second term describes the energy of the $N$ two level emitters  
\begin{equation}    
    H_1 = \hbar\sum_{n=1}^{N}(\Omega_n - i\gamma_n/2)\sigma_n^{+}\sigma_n^{-},\label{AtomicHamiltonian}
\end{equation}
with raising operators $\sigma_n^{+}$ and lowering operators $\sigma_n^{-}$. Here, $\Omega_n$ is the excitation frequency of the $n$-th TLE and $\gamma_n$ represents the decay rate of its excited state via spontaneous emission into modes other than the waveguide continuum. Finally, the last term accounts for the interaction between the TLEs and the photonic modes propagating within the waveguide
\begin{equation}
    H_\text{int} = \frac{\hbar}{2}\sqrt{\frac{\Gamma}{\pi}}\sum_{n=1}^{N}\int d\omega \left[\left(\ell_{\omega}^\dagger e^{i\omega \tau_n} + r_{\omega}^\dagger e^{-i\omega \tau_n}\right)\sigma_n^{-} + \text{h.c.}\right],\label{InteractionHamiltonian}
\end{equation}
where $\Gamma$ is the decay rate of the TLE excitations into waveguide modes and $\tau_n$ represents the propagation time from the first TLE to the $n$-th with a velocity equal to the group velocity of the photonic modes. This Hamiltonian model has been extensively studied in various contexts \cite{Baranger2013,Fan2007,Cirac2015}.

Here we will consider few-photon states scattered from the TLE array. In order to describe the system response we will begin with deriving the transmission and reflection coefficients of a single photon from propagating in the waveguide with a single TLE. We do this using the input-output formalism \cite{InputOutput,Shen2010} of quantum optics. 

For the transmission coefficient $T_1(\omega_k)$ we need to calculate the S-matrix amplitude $\langle k_\text{\tiny R}\vert S\vert p_\text{\tiny R}\rangle = T_1(\omega_k)\delta(\omega_k-\omega_p)$, where we have abbreviated the frequencies by their labels. From now on we will also abbreviate expectation values in the vacuum by omitting the zeroes. By defining the input and output operators as
\begin{align}
    r_{\text{in}}(\omega) &= e^{i H t_0/\hbar}e^{-i H_0 t_0/\hbar}r_\omega e^{i H_0 t_0/\hbar}e^{-i H t_0/\hbar}\label{InputDef}\\
    r_{\text{out}}(\omega) &= e^{i H t_1/\hbar}e^{-i H_0 t_1/\hbar}r_\omega e^{i H_0 t_1/\hbar}e^{-i H t_1/\hbar}\label{OutputDef}
\end{align}
and leveraging the fact that the vacuum is an eigenstate of the full Hamiltonian with eigenvalue zero we can write the scattering amplitude as 
\begin{equation}
    \langle k_\text{\tiny R}\vert S\vert p_\text{\tiny R}\rangle = \langle r_\text{out}(\omega_k)r_\text{in}^{\dagger}(\omega_p)\rangle
\end{equation}
where we have omitted the limit to lighten the notation. By manipulating the Heisenberg picture equation of motion for the right moving photonic modes we can derive the so called input-output relations \cite{InputOutputReviewCirac}
\begin{equation}\label{RInputOutputRel}
    r_\text{out}(\omega) = r_\text{in}(\omega) - i\sqrt{\frac{\Gamma}{2}}\sigma^-(\omega).
\end{equation}
The exact same relation holds for the input-output left modes as well and we have used the notation $\sigma_n^-(\omega)$ for the Fourier transform of the Heisenberg picture time dependent operators $\sigma_n(t)$
\begin{equation}
    \sigma_n^-(\omega) = \frac{1}{\sqrt{2\pi}}\int dt\ e^{i\omega t}\sigma_n^-(t).
\end{equation}
Substituting Eq.~\eqref{RInputOutputRel} and using the fact that the input operators have the usual bosonic algebra, $[r_\text{in}(\omega_1),r_\text{in}^{\dagger}(\omega_2)] = \delta(\omega_1-\omega_2)$ and all other commutators between right moving modes vanish, we can reduce the calculation of the S-matrix amplitude further to
\begin{equation}\label{SinglePhotonSmatix}
    \langle k_\text{\tiny R}\vert S\vert p_\text{\tiny R}\rangle = \delta(\omega_k-\omega_p)-i\sqrt{\frac{\Gamma}{2}}\langle \sigma^-(\omega_k)r_\text{in}^\dagger(\omega_p)\rangle.
\end{equation}
The expectation values in Eq.~\eqref{SinglePhotonSmatix} can be calculated by sandwiching the equation of motion for $\sigma_n^-(t)$ and then Fourier transforming. This can be done analytically and it yields the closed form expression
\begin{equation}\label{AtomicCoefficient}
    \langle \sigma^-(\omega_k)r_\text{in}^\dagger(\omega_p)\rangle = \frac{\sqrt{\Gamma/2}}{\Delta_k+i\frac{\gamma+\Gamma}{2}}\delta(\omega_k-\omega_p)
\end{equation}
with $\Delta_k=\omega_k-\Omega$ is the TLE-photon detuning. Therefore by substituting Eq.~\eqref{AtomicCoefficient} into Eq.~\eqref{SinglePhotonSmatix} and comparing with the definition, we obtain
\begin{equation}\label{TransmissionCoefficient}
    T_1(\omega_k) = \frac{2\Delta_k+i\gamma}{2\Delta_k+i(\gamma+\Gamma)}
\end{equation}
Similarly we can also obtain the reflection coefficient from the S-matrix element $\langle k_\text{\tiny L}\vert S\vert p_\text{\tiny R}\rangle = R_1(\omega_k)\delta(\omega_k-\omega_p)$ which yields
\begin{equation}\label{ReflectionCoefficient}
    R_1(\omega_k) = -\frac{i\Gamma}{2\Delta_k+i(\gamma+\Gamma)}.
\end{equation}

In order to obtain the transmission and reflection coefficients for $N$ emitters we will employ the \textit{transfer matrix} method \cite{TransferMatrixReview2,TransferMatrixReview1}. The transfer matrix is constructed from two matrices. The first one is the \textit{transmission matrix} and it can be derived from the elements of the S-matrix $S = T_1(\omega) + R_1(\omega)\sigma^x$. The S-matrix relates the modes incoming to the emitter to the modes that are outgoing from the emitter while the transmission matrix is relating the modes from one side of the emitter to the other. By rearranging the system of equations accordingly we obtain 
\begin{equation}\label{TransmissionMatrix}
    \mathcal{T}(q) = e^{q(\sigma^z+i\sigma^y)},
\end{equation}
where $q = R_1(\omega)/T_1(\omega)$ is the reflection-transmission ratio. The second matrix is the \textit{free propagation matrix} 
\begin{equation}\label{PropagationMatrix}
    \mathcal{P}(\phi) = e^{i\phi\sigma^z}
\end{equation}
where $\phi=\omega\tau$ represents the phase a photon would pick up from freely propagating for a time $\tau$. This matrix accounts for the phase that left and right moving amplitudes of the photonic wavefunction pick up while propagating freely between the TLEs. From these two matrices we can obtain the \textit{transfer matrix}
\begin{equation}\label{TransferMatrix}
    \mathcal{M}(\lbrace q_n\rbrace,\lbrace \phi_n\rbrace) = \left[\prod_{n=N}^{2}\mathcal{T}(q_{n})\mathcal{P}(\phi_{n-1})\right]\mathcal{T}(q_1)
\end{equation}
where $q_n$ refers to the reflection-transmission ratio evaluated at emitter $n$, $\phi_{n} = \omega(\tau_{n+1}-\tau_{n})$ and with $\prod_{j=N}^{2}$ we imply the ordered product with declining indices from left to right. Finally the transmission and reflection coefficients for the entire array of $N$ emitters, are obtained from the transfer matrix elements as $T_N(\omega) = \mathcal{M}_{22}^{-1}$ and $R_N(\omega) = \mathcal{M}_{21}\mathcal{M}_{22}^{-1}$ respectively.

\begin{figure}
    \centering
    \includegraphics[width=0.49\textwidth]{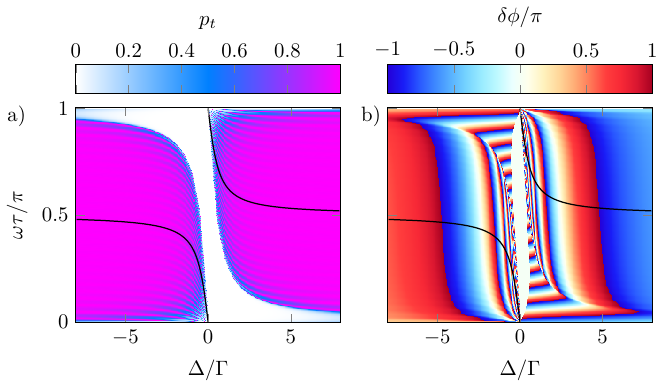}
    \caption{Photon transmission probability (a) and phase shift (b) from a periodic array of $N=30$ TLEs. In both plots the horizontal axes correspond to the ratio of photon-TLE detuning $\Delta$ to the coupling strength $\Gamma$ of Eq.~\eqref{InteractionHamiltonian} while the vertical axes correspond to the propagation phase $\phi = \omega \tau$ that the photon picks up propagating freely from one TLE to the next. Along the black curves the probability of transmission is 1 and the equation describing them is given in Eq.~\eqref{DeterministicTransmissionCurve}. Note that phase shifts are reported in units of $\pi$ and we have assumed $\gamma=0$.}
    \label{fig:Symmetries}
\end{figure}

\section{Results}

\subsection{Phase Gate and State Preparation with a Periodic Array}

In a disorder-free periodic array, Eq.~\eqref{TransmissionCoefficient} allows for closed-form analytic expressions of the transmission and reflection coefficients \cite{NAtomCoefficients1,NAtomCoefficients2} as functions of the emitter-photon detuning $\Delta$ and the phase $\phi = \omega\tau$, where $\tau = \tau_{n+1}-\tau_{n}$ denotes the temporal separation between consecutive emitters. While the explicit expressions are too lengthy to present here, we will visualize them through the transmission probability $p_\text{t} = \vert T_N(\omega)\vert^2$ and phase shift compared to free propagation $\delta\phi = \text{Arg}(T_N(\omega)e^{-i\omega\tau_N})$ in Fig.~\ref{fig:Symmetries}. A similar result was previously reported in \cite{TsoiLaw}.

Building on the results in Fig.~\ref{fig:Symmetries}, we propose a dual-rail photon number qubit encoding protocol \cite{DualRailQubits} to implement a phase gate. In this scheme, each qubit is encoded in two channels. We assign the logical $\vert 0_{\text{L}}\rangle$ to the physical scenario where channel A has one photon and channel B has no photons $\vert 1\rangle_{\text{A}}\vert 0\rangle_{\text{B}}$, while the logical $\vert 1_{\text{L}}\rangle$ state corresponds to $\vert 0\rangle_{\text{A}}\vert 1\rangle_{\text{B}}$. We allow only channel B to pass through the TLE array. The $\vert 0_L\rangle$ state remains unaltered upon passing through the gate, as it corresponds to the vacuum state of the Hamiltonian in Eq.~\eqref{FullHamiltonian}. According to the results presented in Fig.~\ref{fig:Symmetries}, when $\vert 1_{\text{L}}\rangle$ passes through the gate the output state is $e^{i\delta\phi}\vert 1_{\text{L}}\rangle$ with a probability $p_{\text{t}}$ or $\vert 0\rangle_{\text{A}}\vert 0\rangle_{\text{B}}$, which would be a leakage state, with a probability $1-p_{\text{t}}$. Both the phase shift $\delta\phi$ as well as the probability of success of the gate $p_{\text{t}}$ depend on the TLE-photon detuning $\Delta$ as well as the temporal distance of the emitters $\tau$.

We are interested in finding regions of the parameter space $(\Delta/\Gamma, \phi)$ for which the probability of success of our gate $p_\text{t}$ is 1 and simultaneously cover the full range of possible phase shifts $\delta\phi\in\left[-\pi,\pi\right]$. In principle, we can approach this by solving $\vert T_N(\omega)\vert^2=1$ and solve for $\Delta/\Gamma$. In general this is a polynomial equation in $\Delta/\Gamma$ of degree $2N-2$ with $N-1$ double roots, where the polynomial coefficients depend on $\phi$ and $\gamma/\Gamma$. For the $N=30$ case in Fig.~\ref{fig:Symmetries}a one can see these 29 solutions in the form of 29 branches of high probability. Finding these solutions for arbitrary $N$ is not a trivial task, however there is one solution 
\begin{equation}\label{DeterministicTransmissionCurve}
    \Delta = -\frac{\Gamma}{2}\tan\phi
\end{equation}
valid for any even $N$ and for the lossless case $\gamma=0$. We can verify this by substituting $q_n\rightarrow -i\cot\phi$ and $\phi_n\rightarrow\phi$ into Eq.~\eqref{TransferMatrix} and using induction to prove that the transfer matrix is $\mathcal{M}(\phi) = (-1)^{N/2}\mathcal{P}(-\phi)$, which in turn yields the transmission coefficient $T_N(\phi) = (-1)^{N/2}e^{i \phi}$ and therefore the phase shift 
\begin{equation}\label{DeterministicTransmissionPhaseshift}
    \delta\phi = \text{Arg}\left[e^{i\frac{N}{2}\left(\pi-2\phi\right)}\right].
\end{equation}
We note that the argument function wraps the phase in the interval $\left[-\pi,\pi\right]$ therefore the phase of Eq.~\eqref{DeterministicTransmissionPhaseshift} has multiple branches. A branch covers the entire range of phase shifts effectively, if its support is $\left[-\pi,\pi\right]$ in a finite domain and it does not cross the $\Delta =0$. We find that such a branch exists for at least $N = 8$. 

By operating the system along the curve of Eq.~\eqref{DeterministicTransmissionCurve} and by fixing the desired phase shift $\delta\phi$ from Eq.~\eqref{DeterministicTransmissionPhaseshift}, assuming the parameters $\Omega$ and $\Gamma$ appropriately fixed for any particular setup, one obtains the required photon frequency and TLE temporal spacing $\tau$ for the target gate. In the scenario where the frequency $\Omega$ could be somehow controlled, like for superconducting qubits, one could fix the photon frequency and manipulate the phase shift induced by the medium by tuning $\Omega$ and $\tau$ in appropriate values dictated by Eqs.~\eqref{DeterministicTransmissionCurve} and \eqref{DeterministicTransmissionPhaseshift}. Therefore within the same circuit we could induce different phase shifts by applying the gate only as many times as the different phase shifts. For our case however we will assume $\Omega$ fixed and therefore the only valid approach is to fix the photon frequency and TLE spacing $\tau$ such that the phase gate induces an irrational multiple of $\pi$, so that we can densely cover the full range of possible phase shifts by concatenating the gate enough times. This in conjunction with a beam splitter of variable transmission and reflection coefficients would allow the preparation of any possible single qubit state
\begin{equation}
    \vert\psi_q\rangle = \cos\left(\theta/2\right)\vert 0_\text{L}\rangle + e^{i\delta\phi}\sin\left(\theta/2\right)\vert 1_\text{L}\rangle,
\end{equation}
and therefore one would only require a two-qubit operation to define a universal gate set. 

Note that the emitter array can also potentially cover the role of a beam splitter by tuning the parameters appropriately. We assume that we always start with a single photon and a single channel. In the lossless case, the photon will be either transmitted with a probability $p_\text{t} = \vert T_N(\omega)\vert^2$ and a phase $e^{i\delta\phi_\text{t}}$ or it will be reflected with a probability $1-p_\text{t} = \vert R_N(\omega)\vert^2$ and a phase $e^{i\delta\phi_\text{r}}$. Leading the reflected photons to channel A and the transmitted ones to B we can prepare the state
\begin{equation}
    \vert\psi_q\rangle = \sqrt{1-p_\text{t}}\vert 0_\text{L}\rangle + e^{i(\delta\phi_\text{t}-\delta\phi_\text{r})}\sqrt{p_\text{t}}\vert 1_\text{L}\rangle,
\end{equation}
where we have exploited the fact that the state is invariant under multiplication by a global phase. However it is hard to prove rigorously that one can span the entire Bloch sphere for an arbitrary number of atoms and unspecified parameter values.
 
\subsection{Decoherence Effects}

\begin{figure}
    \centering
    \includegraphics[width=0.5\textwidth]{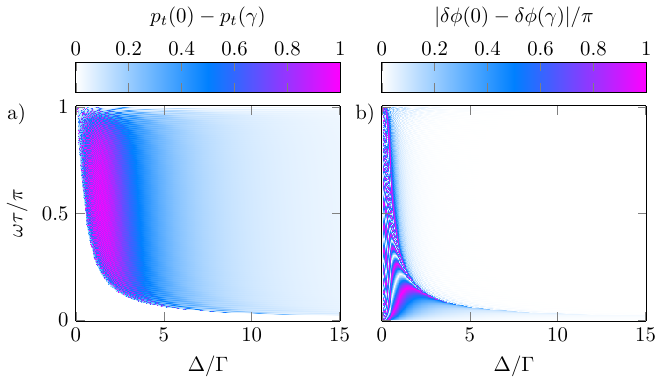}
    \caption{Deviations from the lossless case ($\gamma = 0$), for probability of transmission (a) and phase shift (b) of a photon from a periodic array of 100 TLEs with a finite decay rate to external modes $\gamma$. In both plots the vertical axis correspond to the propagation phase $\phi = \omega \tau$ that the photon picks up propagating freely from one TLE to the next while the horizontal axis correspond to the ratio of the photon-TLE detuning $\Delta$ over the coupling strength $\Gamma$. The value of the loss rate is set at $\gamma = 0.18\Gamma$, which corresponds to a Purcell factor of $\approx$ 35, which is the highest reported value in \cite{StuttgartDevice}.}
    \label{fig:DecoherenceEffects}
\end{figure}

So far we have kept the discussion general by allowing for the possibility of TLE excitation decay into external modes. However in all our plots we have only considered the lossless case of $\gamma=0$ in describing the scattering from the TLE array. Before examining the effects of disorder, it is crucial to first assess the impact of decoherence on our protocol. 

Intuitively, a higher probability of photon absorption by a TLE leads to a greater likelihood of this excitation decaying into external modes. This effect is most pronounced near resonance $\Delta \approx 0$ where the probability of transmission is lowest. This is evident from the fact that the $q$ factor is diverging at resonance for $\gamma=0$. Additionally, in the limit $\tau\rightarrow 0$, the transfer matrix simplifies to $\mathcal{M}(q) = \mathcal{T}(q)^N = \mathcal{T}(Nq)$, meaning that the transmission coefficient resembles that of a single TLE with a rescaled linewidth $\Gamma\rightarrow N\Gamma$. Therefore, for $\tau\rightarrow 0$ and $\Delta/\Gamma \ll N$, we also observe high reflection probability. We refer to this low-transmission region near the origin of the $(\Delta/\Gamma, \omega\tau)$ plot as the \textit{dark} region, while regions of high transmission are termed \textit{bright}. This distinction is illustrated for multiple TLEs in Figs.~\ref{fig:Symmetries}a and \ref{fig:Symmetries}c. 

As shown in Fig.~\ref{fig:DecoherenceEffects}, decoherence effects are most significant near resonance. Specifically, within the bright region, photon loss is more likely near resonance, whereas the TLE spacing plays a lesser role. As expected, in the dark region, transmission probability remains unchanged since photon loss can only reduce transmission. A promising observation is that in the bright region, the phase shift appears unaffected by $\gamma$, with noticeable deviations only at the dark-bright boundary.

Overall, we conclude that deep within the bright region, the system's response is robust against decoherence. For a small number of TLEs, our phase gate protocol covers the full range of phase shifts $\left[-\pi,\pi\right]$ within a detuning range near resonance. For example, at $N=10$, full phase coverage extends approximately from $\Delta\approx 0.4\Gamma$ to $\Delta\approx 1.5\Gamma$. As $N$ increases, the coverage of phase shifts becomes denser, forming multiple fully covered branches, some of which are far from resonance. For instance, at $N=100$, the farthest branch from resonance that still covers the full range of phase shifts lies in the range $\Delta\approx 5\Gamma$ to $\Delta\approx 15\Gamma$. Intuitively, increasing the number of TLEs also raises the likelihood of photon absorption, shifting the "safe operation" regime of our protocol to higher detunings. We estimate this shift to scale like $\sim \gamma N^{2/3}$.

\subsection{Disorder Effects}

\begin{figure*}
    \centering
    \includegraphics[width=\textwidth]{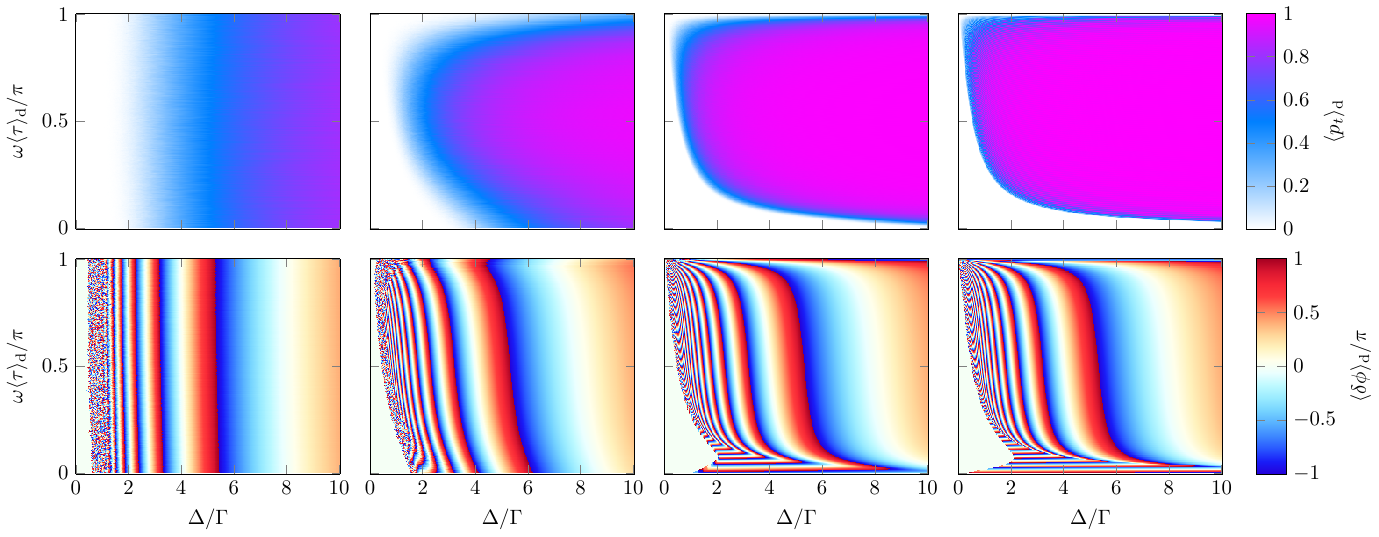}
    \caption{Photon transmission probability and phase shift data from a gas. For each plot the axes are $\Delta/\Gamma$ (horizontal) and average propagation phase $\omega\tau$ (vertical). All plots correspond to a system of $N=100$ TLEs. At each row disorder strength is from left to right, $\frac{\pi}{2},\frac{\pi}{8},\frac{\pi}{32}$ and $\frac{\pi}{128}$. In the first row we present the average transmission probability while in the second the average photon phase shift of Eq.~\eqref{DisroderAveragePhaseShift} in units of $\pi$. Each data point is obtained from averaging over 1000 TLE configurations and we have set $\gamma=0$.}
    \label{fig:GasAnalysis}
\end{figure*}

Our initial motivation for studying the Hamiltonian in Eqs.~\eqref{WaveguideHamiltonian}-\eqref{InteractionHamiltonian} stems from its role as an effective model for the atomic vapor system described in \cite{StuttgartDevice}. Since the atoms in this device form a gas, we must relax the periodicity assumption we have enforced so far to simplify our model and capture key features of the system. In a gas, atomic positions fluctuate between experimental shots. However, assuming the speed of photons in the waveguide is much greater than the thermal velocities of the atoms, we can approximate each shot as a static configuration where atoms occupy fixed but random positions, varying from shot to shot. 

The transfer matrix approach can be directly extended to incorporate disorder. By substituting random phases $\phi_j$ in the propagation matrices of Eq.~\eqref{TransferMatrix} drawn from a Gaussian distribution with a standard deviation $\sigma$ that will play the role of the \textit{disorder strength}, we can compute the transmission probabilities and phase shifts for multiple TLE configurations and averaged over realizations \cite{ChiralWaveguideDisorder}. The results of this calculation are presented in Fig.~\ref{fig:GasAnalysis} and will be discussed here. 

Comparing this to the periodic case (see Fig.\ref{fig:Symmetries}), we observe that both transmission probability and phase shift exhibit weak dependence on TLE spacing for large $N$ and away from resonance. In the bright region, the outgoing photons' phase shift oscillates primarily as a function of $\Delta/\Gamma$, with oscillation frequency decreasing as detuning increases. Consequently, introducing position disorder—drawing adjacent distances from a Gaussian distribution—does not significantly alter the system's response but rather smooths out its weak dependence on $\tau$, as seen in Fig.\ref{fig:GasAnalysis}.

\begin{figure*}
    \centering
    \includegraphics[width=\textwidth]{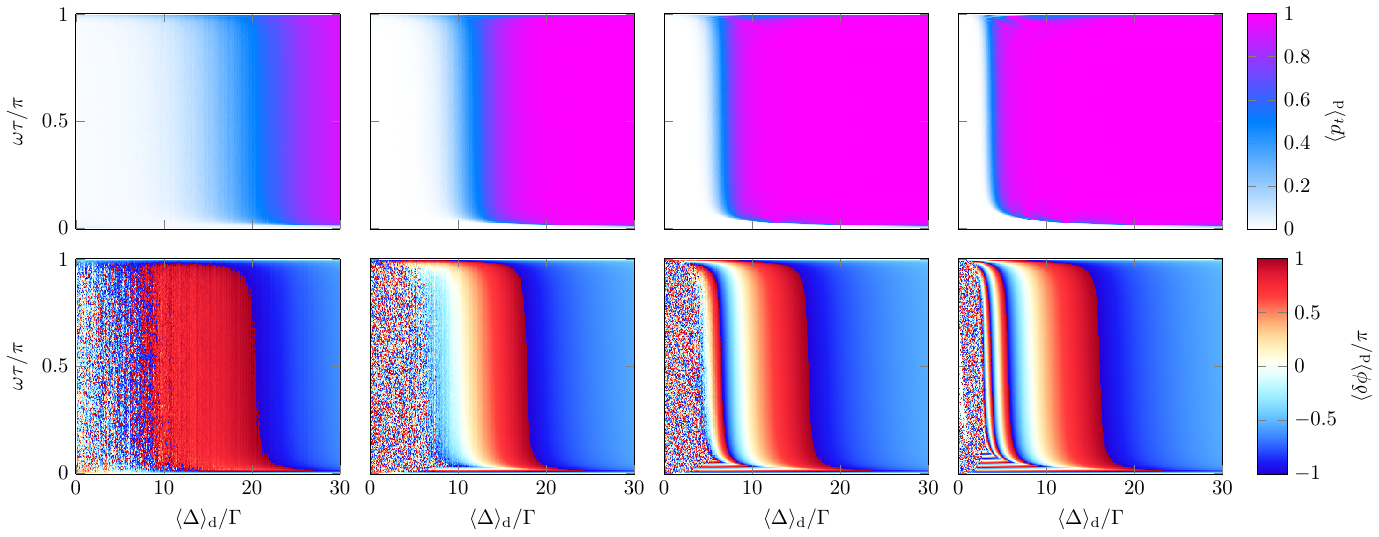}
    \caption{Photon transmission probability and phase shift from a periodic array with TLE frequency disorder. For each plot the axes are average detuning over TLE-waveguide coupling $\langle\Delta\rangle/\Gamma$ (horizontal) and propagation phase $\omega\tau$ (vertical). All plots correspond to a system of $N=100$ TLEs. At each row disorder strength is from left to right, $10\Gamma, 5\Gamma, 2.5\Gamma$ and $ 1.25\Gamma$. In the first row we present the average transmission probability while in the second the average photon phase shift of Eq.~\eqref{DisroderAveragePhaseShift} in units of $\pi$. We note that the noisy parts of the plot are not a convergence issue, but rather an numerical artifact because we are averaging the phases of complex numbers with amplitudes close to numerical precision. This is evident from the fact that the noisy regions of the phase shift plots coincide with the $\sim 0$ probability of transmission regions. Each data point is obtain from averaging over 1000 TLE configurations and we have set $\gamma=0$. We have also obtained the same plots for 10000 realizations per data point, without any discernible improvement for the noisy areas near resonance.}
    \label{fig:DisorderedPeriodicArray}
\end{figure*}

Another symmetry in our initial treatment that we can relax is the assumption of identical transition frequencies for all TLEs. In an atomic vapor setup, frequency disorder can arise due to Doppler shifts from relative atomic motion or the presence of impurities. Similar effects can also emerge in optically trapped atoms due to inhomogeneous potentials or imperfect beam focusing \cite{IonTrapDisorder1,IonTrapDisorder2}.

To isolate the effects of frequency disorder from those of position disorder, we return to the periodic array case and introduce randomness in the TLE transition frequencies, sampling them from a Gaussian distribution. Unlike position disorder, frequency disorder has a pronounced impact on the system's response. Increasing the disorder strength shifts the dark-bright boundary to higher detunings. This result is expected, as a greater disorder strength increases the likelihood that at least one of the $N$ TLEs will be resonant with the incoming photon, and act as a bottleneck for transmission. For sufficiently strong disorder, the system's response becomes nearly uniform, imparting an almost constant phase shift across the bright region regardless of detuning or emitter spacing. Even at lower disorder levels, significant deviations from the structured system response emerge, as illustrated in Fig.~\ref{fig:DisorderedPeriodicArray}. Therefore we conclude that in order to operate the system as a phase gate TLE frequency disorder must be maintained to a minimum.

\subsection{Finite Pulse Scattering and Two-Photon Scattering}

In this section we relax another idealization that we have used, namely that we can prepare the incoming photon to an eigenstate of the system. Since these states are plane waves, in an experimental setup where the light source is turned on at a specific moment the generated light is going to have a finite frequency bandwidth rather than being a pure monochromatic plane wave. Therefore here we will consider an incoming right moving single photon state of the form
\begin{equation}
    \vert\psi_0\rangle = \int dx\ \psi_0(\omega)r_\omega^{\dagger}\vert 0\rangle
\end{equation}
with a Gaussian wavefunction having a bandwidth $\Delta\omega$, localized around the frequency $\omega_c$
\begin{equation}\label{InitialStateWavefunction}
    \psi_0(\omega,\omega_c) = (2\pi\Delta\omega^2)^{-1/4}e^{-\frac{(\omega-\omega_c)^2}{4\Delta\omega^2}}.
\end{equation}

\begin{figure}
    \centering
    \includegraphics[width=0.48\textwidth]{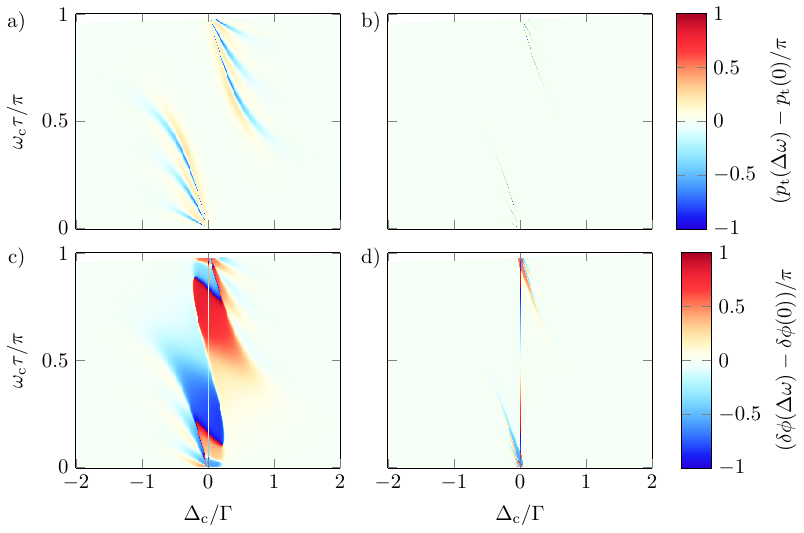}
    \caption{Deviations of probability of transmission (a and b) and phase shift (c and d) for a scattered single photon pulse from the monochromatic case. Panels a) and c) represent data for a broad pulse of bandwidth $\Delta\omega = 0.1\Gamma$ while pannels b) and d) represent the same data for a narrower pulse with bandwidth $\Delta\omega = 0.01\Gamma$. The horizontal axes in all plots correspond to the detuning of the central frequency $\omega_\text{c}$ of the Gaussian pulse of Eq.~\eqref{InitialStateWavefunction} from the TLE transition frequency $\Omega$ in units of $\Gamma$ while the verticals correspond to the free propagation phase $\omega_\text{c}\tau$ in units of $\pi$. For plots we have set $N=4$, the TLE frequency at $\Omega = 100\Gamma$ and we have neglected decoherence effects $\gamma=0$.}
    \label{fig:PulseScattering}
\end{figure}

Since we are now dealing with pulse scattering and long-time dynamics, it is more convenient to employ the input-output formalism again. To compute the transmission probability and phase shift as before, we must evaluate the time evolution of the initial state $\vert\psi_0\rangle$ under the time-independent Hamiltonian of Eq.~\eqref{FullHamiltonian}
\begin{equation}\label{SinglePhotonPulseScattering}
    \langle k_\text{\tiny R}\vert S\vert\psi_0\rangle = \int d\omega_p \psi_0(\omega_p,\omega_c)\langle k_\text{\tiny R}\vert S\vert p_\text{\tiny R}\rangle = T_N(\omega_k)\psi_0(\omega_k,\omega_c)
\end{equation} 
The deviations from the monochromatic case are shown in Fig. \ref{fig:PulseScattering}, where we find that for sufficiently small bandwidth $\Delta\omega$, these deviations remain minimal and are confined to a region near resonance.

For the two-photon case we should begin with calculating the S-matrix amplitudes for a single TLE in the waveguide. In principle we would need to calculate the 9 independent amplitudes (3 possible input states, 3 possible output states) however we will present just one here, namely the amplitude of two right moving photons that both get transmitted 
\begin{equation}\label{TwoPhotonAmplitude}
    \langle k_{1\text{\tiny R}}k_{2\text{\tiny R}}\vert S\vert p_{1\text{\tiny R}}p_{2\text{\tiny R}}\rangle = S_\text{el} + S_\text{in}.
\end{equation}
The first component of the amplitude of Eq.~\eqref{TwoPhotonAmplitude} is
\begin{align}\label{ElasticPart}
    S_\text{el} &= T_{k_1}T_{k_2}[\delta(\omega_{k_1}-\omega_{p_1})\delta(\omega_{k_2}-\omega_{p_2})+p_1\leftrightarrow p_2]
\end{align}
and it describes elastic scattering, where photons either retain their individual energies or exchange them. Here we used the abbreviation for the transmission coefficient of a single photon from the TLE $T_k = T_1(\omega_k)$ in order to lighten the notation. We will use the same notation for the reflection coefficient in the following. This part clearly describes uncorrelated transport of the photons through the emitter and therefore cannot generate entanglement. The second term accounts for inelastic scattering
\begin{equation}\label{InelasticPart}
    S_\text{in} = \frac{2}{\pi\Gamma}R_{k_1}R_{k_2}(R_{p_1}+R_{p_2})\delta(\omega_{k_1}+\omega_{k_2}-\omega_{p_1}-\omega_{p_2}).
\end{equation}
Similar results have been reported in \cite{Shen2010}. The last term describes inelastic scattering, since the photons do not preserve their individual energies, but rather the total energy of the photons. Any entangling that the system can therefore induce would be described by the inelastic term.

\begin{figure}
    \centering
    \includegraphics[width=0.5\textwidth]{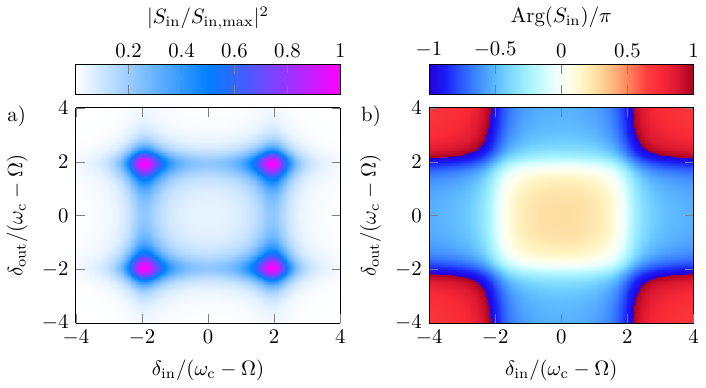}
    \caption{Normalized inelastic amplitude probability density of Eq.~\eqref{InelasticPart} (a) and corresponding phase shift distribution (b). We also use a "center of mass" frame of the two photons where we defined the central frequency $\omega_\text{c}=(\omega_1+\omega_2)/2$ and photon-photon detuning $\delta = \omega_1-\omega_2$. The labels "in" and "out" indicate that the corresponding values are calculated for the input or output photon frequencies respectively. Note that in this reference frame the Dirac delta constraint simply implies $\omega_\text{c,out} = \omega_\text{c,in}$. From these definitions we can read directly from the plots that the inelastic amplitude is relevant only if at least one of the incoming photons is on resonance, since the peaks occur at $\omega_\text{c}\pm\frac{\delta}{2} = \Omega$. In both plots we have set $\Omega = 100\Gamma$, $\gamma = 0$ and we have set the conserved central frequency at $\omega_\text{c} = \Omega+2\Gamma$.}
    \label{fig:TwoPhotonScattering}
\end{figure}

Unfortunately, it is due to the inelastic term that employing a transfer matrix approach is not possible because the input modes couple to a continuum of output modes. Ignoring the inelastic term, we could use the transfer matrix approach however this is trivial uncorrelated transport of two photons through the array, and therefore irrelevant for two qubit gates. Furthermore, obtaining the transmission and reflection coefficients directly by solving the Langevin equations for $N$ emitters is only possible at the weak excitation limit $\sigma^z(t)\approx -1$. However as it can be seen from Fig~\ref{fig:TwoPhotonScattering} the entangling inelastic term is prominent only for resonant input states and therefore we cannot employ the weak excitation limit. Finally, while single-TLE scattering can, in principle, generate entanglement, it is not a reliable mechanism for deterministic entangling operations. The elastic term overwhelmingly dominates, limiting the efficiency of entanglement generation. Increasing the interaction strength is also not a viable option, as stronger interactions typically increase the emitter’s loss to non-waveguide modes.

\section{Discussion}

We have established that a one-dimensional array of two-level emitters (TLEs) can function as a deterministic phase gate for flying photonic qubits using dual-rail encoding. This functionality remains robust even in the presence of dissipation into non-waveguide modes, with the operating regime simply shifting to higher detunings as the number of atoms increases. While our initial analysis assumed a periodic array, we have demonstrated that the phase gate protocol remains effective even when periodicity is relaxed, making it applicable to atomic vapor setups such as the one in \cite{StuttgartDevice}.

However, our findings also highlight a crucial limitation: while the system is resilient to decoherence and positional disorder, it is highly sensitive to frequency disorder among the TLEs. This underscores the importance of precise control over atomic transition frequencies for practical implementations. Additionally, in a more realistic scenario where a photon pulse rather than a monochromatic wave is scattered, we have shown that the phase gate remains accurate for realistic pulse bandwidths.

To establish this platform as a viable approach to quantum computing, it is essential to demonstrate the capability for two-photon interactions—an essential requirement for universal quantum logic. While we have identified nonlinear effects in the transmission of two photons through a single TLE that could facilitate entanglement, our current methodology is not well suited for analyzing cooperative effects in large arrays of emitters. Given that such cooperative phenomena may be the key to enhancing inelastic scattering and generating stronger photonic interactions, future work will focus on extending our analysis to the many-emitter case.

\section{Methods}

\subsection{Photon Loss Scaling}

To quantify how the onset of the safe operation regime for our phase gate protocol shifts with increasing emitter number $N$ and emitter losses $\gamma$ to non-waveguide modes, we performed the following analysis. We defined a rectangular window in the $(\Delta/\Gamma,\phi)$ parameter space, with $\Delta/\Gamma\in[0,20]$ and $\phi\in[0,\pi/2]$, and computed the transmission loss across this window for various values of $N$ and $\gamma$, as shown in Fig.~\ref{fig:DecoherenceEffects}a.

To extract a single figure of merit from this data, we introduced a threshold transmission loss of 0.1 and calculated the area within the parameter window where the loss exceeds this threshold. We denote the ratio of this region to the total window area as $A_\gamma(N)$. Although the threshold value is arbitrary and does not correspond to an optimized cutoff, we verified that the scaling behavior is largely insensitive to its specific choice. The threshold of 0.1 was chosen to ensure that the ratio $A_\gamma(N)$ remains sensitive to variations in $N$ and $\gamma$ without saturating within the bounds of the selected window.

Notably, as $N$ increases, the affected region in parameter space becomes more localized, and $A_\gamma(N)$ eventually saturates to a value less than unity due to the mismatch between the rectangular window and the actual geometry of the loss region. Nevertheless, for moderate $N$, the ratio exhibits clear scaling behavior. A least-squares fit of the data in log-log space reveals a power-law dependence of the form
\begin{equation}
    A_\gamma(N) \sim \gamma N^{2/3}
\end{equation}
highlighting how both emitter loss and system size influence the detuning required to operate in the high-fidelity regime.

\subsection{Disorder Averaging}

A key consideration when averaging phase shifts is the potential for incorrect results when averaging values algebraically, particularly near $\pm\pi$. To mitigate this, we instead perform a vectorial average in the complex plane before extracting the argument
\begin{equation}\label{DisroderAveragePhaseShift}
    \langle\delta\phi\rangle_\text{d} = \text{Arg}\left(\sum_{r}p_c(r)e^{i\delta\phi(r)}\right)
\end{equation}
where $r$ labels a specific disorder realization, $p_{\text{c}}(r)$ is the probability of realization $r$ and the $\delta\phi(r)$ is the phase shift induced by realization $r$.

A final remark regarding the statistical treatment of randomly positioned TLEs concerns our definition of the mean. We modeled the spacings between adjacent TLEs using a Gaussian distribution. However, as seen in Fig.~\ref{fig:GasAnalysis}, we specifically considered a scenario where $\langle\phi\rangle_\text{d} = \omega\langle \tau\rangle_\text{d} = 0$ with a finite standard deviation. This setup would, in principle, include negative spacings, which are unphysical. To address this, we discarded negative $\phi_j$ values. While this alters the actual distribution's mean compared to the original Gaussian, we conventionally denote $\langle \phi\rangle_\text{d}$ as the mean before applying this constraint.

\subsection{Pulse Scattering Probability and Phase Shift}

For finite-bandwidth pulse scattering, the transmission probability and phase shift can no longer be extracted directly from the transmission coefficient alone. The spatiotemporal profile of the pulse induces frequency-dependent variations in both observables, necessitating a frequency-averaging procedure. Based on the output state amplitudes in Eq.~\eqref{SinglePhotonPulseScattering}, we define the generalized transmission probability for a pulse centered at $\omega_c$ as
\begin{equation}\label{PulseTransmissionProbability} p_{\text{t}}(\omega_c) = \int d\omega_k\ \vert T_N(\omega_k)\vert^2\vert\psi_0(\omega_k,\omega_c)\vert^2, \end{equation}
and the corresponding phase shift as
\begin{equation}\label{SpaceAveragedPhaseShift} \delta\phi(\omega_c) = \text{Arg}\left(\int d\omega_k\ T_N(\omega_k)\vert T_N(\omega_k)\vert^2\vert\psi_0(\omega_k,\omega_c)\vert^2\right). \end{equation}
Here, we employ the same vectorial averaging technique used in the disorder analysis, which ensures a consistent treatment of the complex phase and avoids artifacts due to phase wrapping.

\section{Data Availability}

Data is available from the corresponding author upon reasonable request.

\section{Code availability}

The codes used to generate data for this paper are available from the corresponding author upon reasonable request.

\section{Acknowledgments} 

J.A. and E.V thank R. Löw and T. Pfau for their valuable input. E.V. would like to thank Konstantinos Rafail Revis for the fruitful discussions. We gratefully acknowledge financial support from the Carl-Zeiss-Foundation via the Carl-Zeiss-Center for Quantum Photonics (QPhoton) and the Competence Center for Quantum Computing Baden-Württemberg (KQCBW). 

\section{Competing interests}

The authors declare no competing interests.

\section{Contributions}

Both authors contributed to verifying the results and writing the manuscript. J.A. conceived the project and supervised the work. E.V. developed the phase gate protocol and performed the numerical and analytic calculations presented in the paper.

\bibliography{npjReferences.bib}

\end{document}